\documentclass[fleqn,useAMS,usenatbib]{mn2e}
\usepackage{color}
\usepackage{exscale}
\usepackage{graphicx}
\usepackage{natbib}
\usepackage{rotating}
\usepackage{rotate}
\usepackage{afterpage}
\usepackage{fancyhdr}	
\usepackage{amsmath}
\usepackage{amssymb}
\usepackage{relsize}
\usepackage{lscape}
\usepackage{epsfig}
\usepackage{tabularx}
\usepackage{longtable}
\usepackage{ltxtable}
\usepackage{txfonts}
\usepackage{natbib}
\usepackage{bm}

\def\aap{A\&A}
\def\apj{ApJ}
\def\apjl{ApJL}

\def\mnras{MNRAS}

\def\physrep{Phys.~Rep.}   
\def\prd{Phys.~Rev.~D}

\newcommand{\tbf}{\textbf}
\newcommand{\ti}{\textit}

\newcommand{\bea}{\begin{eqnarray}}
\newcommand{\be}{ \begin{equation}}
\newcommand{\ben}{\begin{enumerate}}
\newcommand{\bi}{\begin{itemize}}
\newcommand{\eea}{\end{eqnarray}}
\newcommand{\ee}{ \end{equation}  }
\newcommand{\ei}{\end{itemize}}
\newcommand{\een}{\end{enumerate}}
\newcommand{\nn}{\nonumber}

\newcommand{\matC}{\mathbf C}

\newcommand{\om}{\Omega_\mr m}
\newcommand{\omb}{\Omega_\mr b}
\newcommand{\sig}{\sigma_8}

\newcommand{\vt}{\vartheta}

\newcommand{\eps}{\epsilon}

\newcommand{\mr}{\mathrm}

\renewcommand{\d}{{\rm d}}

\def\tmin{{\vt_{\rm min}}}
\def\tmax{{\vt_{\rm max}}}

\def\xipm{\xi^{\gamma \gamma}_\pm}
\def\xip{\xi^{\gamma \gamma}_+}
\def\xim{\xi^{\gamma \gamma}_-}
\def\ximm{\xi^{\mu \mu}} 
\def\xipp{\xi^{g g }} 
\def\xism{\xi^{\gamma \mu}} 
\def\xisp{\xi^{\gamma g}} 
\def\ximp{\xi^{\mu g}} 

\def\wss{W^{\gamma \gamma}}
\def\wmm{W^{\mu \mu}}
\def\wpp{W^{g g}}
\def\wsm{W^{\gamma \mu}}
\def\wsp{W^{\gamma g}}
\def\wmp{W^{\mu g}}

\def\tss{T^{\gamma \gamma}}

\def\css{C^{\gamma \gamma}}
\def\cmm{C^{\mu \mu}}
\def\cpp{C^{g g}}
\def\csm{C^{\gamma \mu}}
\def\csp{C^{\gamma g}}
\def\cmp{C^{\mu g}}

\def\ess{\vek E^{\gamma \gamma}}

\def\emm{\vek E^{\mu \mu}}
\def\epp{\vek E^{g g}}
\def\esm{\vek E^{\gamma \mu}}
\def\esp{\vek E^{\gamma g}}
\def\emp{\vek E^{\mu g}}

\def\covssss{\tbf{C}^{\gamma \gamma \gamma \gamma}}
\def\covmmmm{\tbf{C}^{\mu \mu \mu \mu}}
\def\covpppp{\tbf{C}^{ g g g g }}
\def\covsmsm{\tbf{C}^{\gamma \mu \gamma \mu}}
\def\covspsp{\tbf{C}^{\gamma g \gamma g}}
\def\covpmpm{\tbf{C}^{g \mu g \mu}}

\def\covssmm{\tbf{C}^{\gamma \gamma \mu \mu}}
\def\covsspp{\tbf{C}^{\gamma \gamma  g g}}
\def\covsssm{\tbf{C}^{\gamma \gamma \gamma \mu}}
\def\covsssp{\tbf{C}^{\gamma \gamma \gamma g}}
\def\covsspm{\tbf{C}^{\gamma \gamma g \mu}}

\def\covppmm{\tbf{C}^{g g \mu \mu}}
\def\covmmsm{\tbf{C}^{\mu \mu \gamma \mu}}
\def\covspmm{\tbf{C}^{\gamma g \mu \mu}}
\def\covmmpm{\tbf{C}^{\mu \mu g \mu}}

\def\covppsm{\tbf{C}^{g g \gamma \mu}}
\def\covsppp{\tbf{C}^{\gamma g g g}}
\def\covpppm{\tbf{C}^{g g g \mu}}

\def\covspsm{\tbf{C}^{\gamma g \gamma \mu}}
\def\covsmpm{\tbf{C}^{\gamma \mu g \mu}}

\def\covsppm{\tbf{C}^{\gamma g g \mu}}

\def\tssss{T^{\gamma \gamma \gamma \gamma}}
\def\tpppp{T^{gggg}}
\def\tdddd{T^{\delta \delta \delta \delta}}

\newcommand{\like}{L}
\newcommand{\prob}{P}
\newcommand{\pco}{\vek p_\mr{co}}

\def\vek{\mathbf}

\title[Combining Probes of Large-Scale Structure with \textsc{CosmoLike}]{Combining Probes of Large-Scale Structure with \textsc{CosmoLike}}
\author[Eifler et al.]
{\parbox{\textwidth}{Tim Eifler$^{1,2}$\thanks{E-mail: \texttt{timeifler@gmail.com}},
Elisabeth Krause$^{1}$,
Peter Schneider$^{3}$,
Klaus Honscheid$^{2,4}$ \vspace{0.4cm}}\\
\parbox{\textwidth}{$^{1}$ Department of Physics and Astronomy, University of Pennsylvania, Philadelphia, PA 19104, USA \\
$^{2}$ Center for Cosmology and Astro-Particle Physics, The Ohio State University,  Columbus, OH 43210, USA\\
$^{3}$ Argelander Institut f\"ur Astronomie, Universit\"at Bonn,  53121 Bonn, Germany\\
$^{4}$ Department of Physics, The Ohio State University, Columbus, OH 43210, USA\\
}}

\begin{document}

\date{accepted received}

\pagerange{\pageref{firstpage}--\pageref{lastpage}} \pubyear{2010}

\maketitle

\label{firstpage}

\begin{abstract}
{Developing accurate analysis techniques to combine various probes of cosmology is essential to tighten constraints on cosmological parameters and to check for inconsistencies in our model of the Universe.\\
In this paper we develop a joint analysis framework for six different second-order statistics calculated from three tracers of the dark matter density field, namely galaxy position, shear, and magnification. We extend a data compression scheme developed in the context of shear-shear statistics (the so-called COSEBIs) to the other five second-order statistics, thereby significantly reducing the number of data points in the joint data vector.\\
We use \textsc{CosmoLike}, a newly developed software framework for joint likelihood analyses, to forecast parameter constraints for the Dark Energy Survey (DES). The simulated MCMCs cover a five dimensional cosmological parameter space comparing the information content of the individual probes to several combined probes (CP) data vectors. Given the significant correlations of these second-order statistics we model all cross terms in the covariance matrix; furthermore we go beyond the Gaussian covariance approximation and use the halo model to include higher order correlations of the density field.\\   
We find that adding magnification information (including cross probes with shear and clustering) noticeably increases the information content and that the correct modeling of the covariance (i.e., accounting for non-Gaussianity and cross terms) is essential for accurate likelihood contours from the CP data vector.\\
We also identify several nulltests based on the degeneracy of magnification and shear statistics which can be used to quantify the contamination of data sets by astrophysical systematics and/or calibration issues.}
\end{abstract}


\section{Introduction}
\label{sec:intro}
High quality data sets from near-term wide-field imaging surveys, e.g. Kilo-Degree Survey (KiDS\footnote{http://www.astro-wise.org/projects/KIDS/}), Hyper Suprime Cam (HSC\footnote{http://www.naoj.org/Projects/HSC/HSCProject.html}), Dark Energy Survey (DES\footnote{www.darkenergysurvey.org/}), allow for tight constraints on cosmological parameters from the Large Scale Structure (LSS) of the Universe, being complementary with Cosmic Microwave Background (CMB) constraints from the Wilkinson Microwave Anisotropy Probe (WMAP\footnote{http://map.gsfc.nasa.gov/}) \citep[see][and references therein]{glk12} and Planck\footnote{http://www.esa.int/Our\textunderscore Activities/Space\textunderscore Science/Planck}. The improved data quality and the small statistical uncertainties (as a result of the increased survey volume) pose new challenges for the data analysis; the development and refinement of LSS data analysis methods is crucial for the success of even larger, future data sets from the Large Synoptic Survey Telescope (LSST\footnote{http://www.lsst.org/lsst}), and from future satellite missions Euclid\footnote{sci.esa.int/euclid/} \citep{laa11} and the Wide-Field Infrared Survey Telescope (WFIRST\footnote{http://wfirst.gsfc.nasa.gov/}).\\  
For all of the aforementioned surveys the tightest constraints on cosmology will be obtained from a joint analysis of all probes that can be extracted from the data (e.g., cluster mass function, shear peak statistics, BAO peak fitting, and various second-order statistics derived from clustering, shear, and magnification). Combining LSS with Supernovae and CMB constraints is straightforward; due to the fact that these probes have very little correlation a joint likelihood analysis frequently comes down to multiplying the corresponding posteriors probabilities \citep[e.g.,][]{kfh12}. Combining the various probes of LSS themselves is complicated for several reasons: First, the cosmological information of various LSS probes is highly correlated, which prohibits a joint analysis on the level of posterior probabilities. Instead the analysis requires a joint likelihood using a covariance matrix that includes all cross correlation terms between the individual probes. Second, not only is the cosmological information correlated, even more problematic are the correlations of various systematic effects originating from astrophysics and the measurements themselves. \\ 
\textsc{CosmoLike}, a new analysis framework for high accuracy CP analyses, includes the covariance matrix's cross terms in the likelihood analysis, moreover it consistently models the CP model vector as a function of cosmology and also as a function of the uncertainties in the nuisance parameters. Developing such a CP prediction code is challenging given that modeling the individual probes already requires refined knowledge and high-level expertise on the corresponding astrophysics and systematics. Although this knowledge is present in the corresponding communities, even the individual analysis methods are under constant development in order to meet the new data quality, and unfortunately these methods are largely independent from each other. Phrasing the problem differently: the large correlation of the LSS probes is not reflected in the correlation of the development of the individual analysis techniques.\\
For example, probably the most important astrophysical uncertainty for clustering based measurements is the relation of dark and luminous matter, modeled through various bias parametrizations and/or Halo Occupation Distribution (HOD) models. Constraints on these models come from measuring cross correlations of shear and clustering (sometimes called galaxy-galaxy lensing). Cosmic shear uses the same cross terms to offset uncertainties due to intrinsic alignment; simply combining both methods uses the galaxy-galaxy lensing information twice. Similar problems occur when modeling shear calibration which affects cluster masses calibrated through weak lensing, cosmic shear, galaxy-galaxy lensing, and shear peak statistics, all at the same time. Other examples are the modeling of baryonic uncertainties and photo-z calibration which affect all probes but in different ways.\\  
Whereas solving all these problems is beyond the scope of this paper, it is our intention nevertheless to take first steps towards a coherent analysis framework of LSS probes. We limit our problem to second-order statistics (power spectra, correlation functions or linear transformations thereof) that can be derived from a galaxy catalog containing measurements of galaxy shear, galaxy position and galaxy magnification. As we explain further in Sect. \ref{sec:concepts} we obtain six different second-order statistics from these measurements corresponding to six different probes of the density field. We exclude galaxy clusters and shear peak statistics for now, since these are first-order number count statistics which cause additional complications in the sense that they require a different likelihood function (Poisson distribution instead of Gaussian) in their analysis. Incorporating these probes at the level of the covariance matrix in a Gaussian likelihood together with second-order statistics is questionable.\\
Regarding nuisance parameters we account for uncertainties from modeling bias and correlation parameters that affect all probes involving clustering (see Sect. \ref{sec:bias}), however the extension to other astrophysical contaminations, e.g. intrinsic alignment and baryonic effects is straightforward, and although not being part of the analysis we address it in the discussion. \\
Within the aforementioned restrictions \textsc{CosmoLike v1.0}, which we use in this paper, advances the existing state of the art of simulated likelihood analysis: 
\begin{enumerate} 
 \item We simulate an actual likelihood analysis in a five dimensional cosmological parameter space (plus ten parameters for modeling bias and correlation parameter)
 \item We use a full Non-Gaussian covariance which includes correlations between all probes and account for higher order correlations of the density field (see Sect. \ref{sec:cov})
 \item We develop a data compression scheme for the joint likelihood analysis which simultaneously solves the shear-shear EB-mode problem.
\end{enumerate}
This data compression scheme was developed by \citet{sek10} (hereafter SEK10) to solve the problem of calculating the a shear E-mode two-point statistics (which contains the cosmological information) from a given shear-shear correlation function on a finite interval. Since information from shear data is limited to angular scales $[\tmin ; \tmax]$ any E/B-mode statistic which requires information on larger or smaller scales suffers from so-called E/B-mode mixing or leakage. This problem is examined in \citet{kse06} for configuration space quantities, finding a significant (scale-dependent) bias for formerly used shear statistics, e.g. aperture mass dispersion or shear dispersion. The issue has been addressed in even greater detail for Fourier space quantities, mostly in the context of CMB polarization experiments; several groups developed and refined a Pseudo-Cl technique \citep{hgn02,bct05} that has been applied to simulated shear data in \citet{hth11}. In Fourier space E/B-leakage largely depends on the mask of the survey; several mitigation schemes have been developed \citep{lew03,smi06,kin10}. \\
Except for shear-shear none of the other five second-order statistics suffers from the EB-mode problem; nevertheless the data compression aspects of the COSEBIs are highly desirable for these probes as well. Furthermore, the extension of the COSEBIs scheme allows for a joint cosmological analysis that involves a clean separation of the cosmic shear signal into E- and B-modes.

\section{Basic concepts}
\label{sec:concepts}
We consider the observables shear $\gamma$, magnification $\mu$, and galaxy position $g$. From these observables the following second-order statistics can be obtained: shear-shear ($\gamma \gamma$), magnification-magnification ($\mu \mu$), galaxy position-galaxy position ($g g$), shear-magnification ($\gamma \mu$), shear-position ($\gamma g$), and magnification-position ($\mu g$). We want to comprise this second-order cosmological information into a COSEBIs data vector
\be
\label{eq:dv}
\vek E = (\ess, \esp, \epp, \emm,  \esm,  \emp)^\mr t \,,
\ee
where each $\vek E^{xx}$ contains five COSEBI modes \citep[see SEK10,][for justification of the number of modes]{eif11, ass12}. The goal of this paper is to simulate a multi-dimensional likelihood analysis, where ``simulated'' means that $\vek E$ is computed from a fiducial cosmological model (see Table \ref{tab:tab2}) using our prediction code; we will refer to this data vector as the \ti{fiducial data vector} from now on.\\
We assume that the errors of the input data vector $\vek E$ are described by a multivariate Gaussian 
\be
\label{eq:likelihood}
L (\vek E|\pco )=\frac{1}{(2\pi)^{N/2}\sqrt{|\matC|}} \mr{exp} \left[ - \frac{1}{2} \left( \vek E - \vek M \right)^t  \matC^{-1} \left(\vek E - \vek M \right) \right] \;, 
\ee
where $\pco$ denotes the cosmological parameter vector that is assumed in the model vector $\vek M$, hence  $\vek M \equiv \vek M (\pco)$.\\
The posterior probability in cosmological parameter space is obtained via Bayes' theorem
\be
\label{eq:bayes}
\prob(\pco|\vek E) = \frac{\prob (\pco) \,\like (\vek E| \pco )}{\prob (\vek E)} \,,
\ee
where $\prob (\pco )$ denotes the prior probability (we assume non-informative priors) and the evidence $\prob (\vek E)$ can be calculated as an integral over the likelihood $\prob (\vek E)=\int \d \pco \prob (\pco) \, \like(\vek E | \pco)$ providing a normalization constant for the posterior probability.\\
Given the functional form of the likelihood as in Eq. (\ref{eq:likelihood}) the error bars are fully determined by the covariance of the COSEBIs' data vector, which correspondingly to the definition in Eq. (\ref{eq:dv}) reads
\be
\label{eq:covconcept}
\vek C= \left( \begin{array}{c | c | c | c | c | c }
\covssss &  \covsssp & \covsspp & \covssmm & \covsssm & \covsspm\\ \hline
 & \covspsp & \covsppp & \covspmm & \covspsm & \covsppm \\ \hline
 & &\covpppp & \covppmm & \covppsm &\covpppm \\ \hline
 & & & \covmmmm & \covmmsm& \covmmpm \\ \hline
 & & & & \covsmsm& \covsmpm\\ \hline
 & & & & & \covpmpm\\ 
\end{array}\right) \,,
\ee
with $\vek C$ being symmetric. While postponing a detailed description of the covariance's modeling to Sect. \ref{sec:cov}, we note that we assume the covariance to be constant with respect to the point in parameter space where the likelihood is evaluated. As shown in \citet{esh09} (for Gaussian shear-shear covariances) this assumption is problematic and, depending on the survey parameters, can have significant impact on the parameter constraints. We acknowledge that the covariance matrix, since predicted from a cosmological model, in principle has to vary with respect to cosmology \citep[see][for corresponding application to shear data]{kfh13} and we will pursue a corresponding extension of this work in the future. \\
In practice the COSEBIs are calculated from the correlation functions of the three observables. The corresponding power spectra are related to these correlation functions as 
\bea
\label{eq:2PCFshear}
\xipm (\vt) &=& \frac{1}{2 \pi} \int \d l \,  l \, J_{0/4} (l \vt) \, \css (l) \,, \\
\label{eq:2PCFmag}
\ximm (\vt) &=& \frac{1}{2 \pi} \int \d l \,  l \, J_0 (l \vt) \, \cmm (l) \,, \\
\label{eq:2PCFgal}
\xipp (\vt) &=& \frac{1}{2 \pi} \int \d l \,  l \, J_0 (l \vt) \, \cpp (l) \,, \\
\label{eq:2PCFsm}
\xism (\vt) &=& \frac{1}{2 \pi} \int \d l \,  l \, J_2 (l \vt) \, \csm (l)\,, \\
\label{eq:2PCFsp}
\xisp (\vt) &=& \frac{1}{2 \pi} \int \d l \,  l \, J_2 (l \vt) \, \csp (l)\,, \\
\label{eq:2PCFmp}
\ximp (\vt) &=& \frac{1}{2 \pi} \int \d l \,  l \, J_0 (l \vt) \, \cmp (l)\,,
\eea
where we point out the $J_2$ in the polar-scalar correlation functions $\xism$ and $\xisp$ \citep[see e.g.][for a derivation]{bas01}. We will return to the filter functions in Sect. \ref{sec:nulltests}.

\section{Modeling the data vector}
\label{sec:model}
\begin{figure}
\includegraphics[width=9cm]{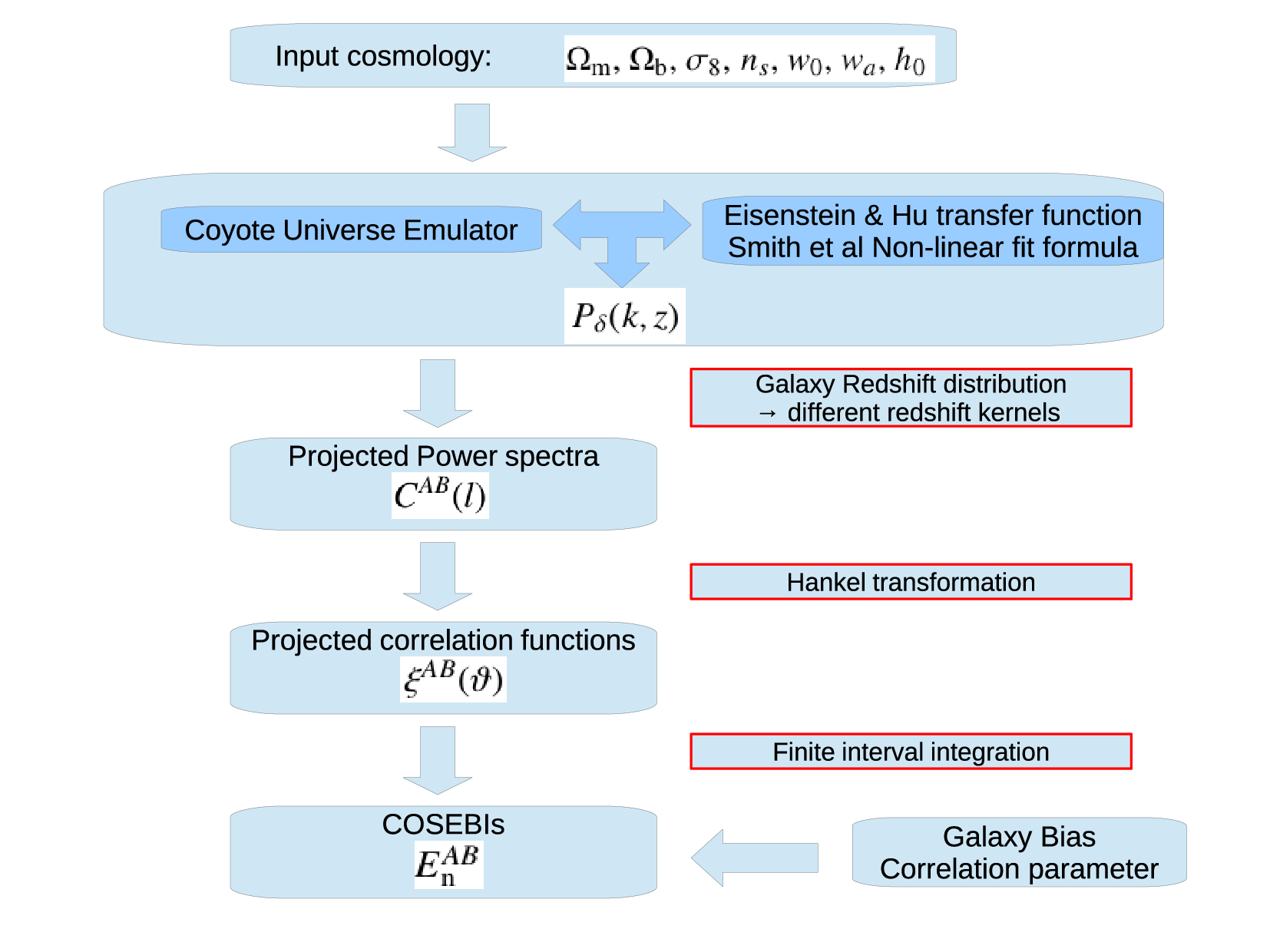}
 \caption{Schematic illustration of the modeling of the CP COSEBIs data vector for a given cosmology.}
         \label{fi:modelscheme}
\end{figure}

In this section we describe the prediction module of \textsc{CosmoLike v1.0} (see Fig. \ref{fi:modelscheme} for an illustration), which is an extended version of the shear-shear prediction code described in \citet{eif11}. All projected quantities are computed from the nonlinear density power spectrum which we calculate from an initial power spectrum $P_{\delta \delta}(k) \propto k^{n_\mr s}$ using the transfer function of \citet{eih98}. In order to model the non-linear evolution of the density field we develop a Hybrid approach combining information from Halofit \citep{smp03} and the Coyote Universe Emulator \citep{lhw10}. The latter emulates $P_{\delta \delta}$ over the range $k \in [0.002;3.4]$ $h$/Mpc within $z \in [0;1]$ to an accuracy of $1\%$ for cosmologies within $\om h^2 \in [0.120;0.155]$, $\omb h^2 \in [0.015;0.0235]$, $n_\mr s \in [0.85;1.05]$, $\sig \in [0.6;0.9]$, $w_0 \in [-1.3;-0.7]$. For any cosmology, $z$, and $k$ within the aforementioned range, we solely rely on the output of the emulator. For all other parameters we compute the non-linear part of $P_{\delta \delta}$ from Halofit and rescale this solution by a factor 
\be
f(k,z,\pco)= \frac{P_{\delta \delta}^\mr{Coyote} (k,z,\pco^\mr{close})}{P_{\delta \delta}^\mr{Halofit} (k,z,\pco^\mr{close})} \,
\ee
where $\pco$ is the cosmology parameter vector of interest and $\pco^\mr{close}$ is the closest point in parameter space where the Emulator returns a solution (close is defined as minimum difference in each parameter separately).\\
In order to simulate wCDM models we follow the strategy outlined in {\sc icosmo} \citep{rak11}, which interpolates Halofit between flat and open cosmological models to mimic Quintessence cosmologies \citep[please also see][for more details]{shj10}. Outside the parameter range of the Emulator the precision of $P_{\delta \delta}$ will of course be significantly below $1\%$; we nevertheless believe that our approach supersedes other implementations of modeling non-linear structure growth for multiple cosmologies. For example, when using Halofit alone it has been shown that the lensing power spectrum is substantially underestimated \citep[e.g.,][]{hhw09}.
Throughout this paper we assume a redshift distribution as expected from DES. More precisely, this is modeled by modifying a redshift distribution measured from the Canada-France-Hawaii Telescope Legacy Survey \citep[see][adjusted for the lower mean redshift of DES]{bhs07}. The exact parameterization reads
\be 
\label{redshiftben}
n(z)=N \, \left( \frac{z}{z_0}\right)^\alpha \exp \left[ - \left(  \frac{z}{z_0} \right)^\beta \right]\,,
\ee 
with $\alpha=1.3$, $\beta=1.5$, $z_0=0.56$.
\subsection{Modeling the projected power spectra}
\label{sec:power spectra}
\cite{kai92,kai98} show that projected power spectra are related to the 3D power spectrum of density fluctuations $P_{\delta \delta}$ via a Fourier equivalent of Limber's equation, i.e.
\be
P_{12} (l)= \int \d \chi \frac{q_1(\chi) \, q_2(\chi)}{f^2_k(\chi)} \, P_{\delta \delta} \left( k,\chi \right) \, \,
\ee
with $q_1, q_2$ being weight functions, $k=l/\chi$, and $f_k(\chi)$ being the comoving angular diameter distance which corresponds to the comoving coordinate $\chi$ for the case of vanishing curvature. For simplicity, we will assume the latter in our analysis; note that the tools described in this paper are nevertheless independent of this assumption. \\
In case of the shear the weight functions $q$ read
\be
q_i=\frac{3 H_0^2 \, \om}{2 c^2} \, \frac{g_i(\chi) \, \chi}{a} = \frac{3 H_0^2 \, \om}{2 c^2} \, \frac{\chi}{a} \int_\chi^{\chi_\mr h} \d \chi' \, p_i(\chi') \, \frac{\chi' - \chi}{\chi'} \,,
\ee
where $a(\chi)$ is the scale factor and $p_i(\chi) \d \chi=p_i(z) \d z$ is the redshift distribution of source galaxies in the $i^\mr{th}$ tomography bin. We do not consider tomography in this paper and drop the corresponding denotation of different redshift bins from now on.\\
Using these weight functions the expression for the shear power spectrum reads
\be
\label{eq:css}
\css (l) =\frac{9}{4} \left( \frac{H_0}{c} \right)^4 \om^2 \int_0^{\chi_h} \d \chi \, \frac{g^2 (\chi) }{a^2(\chi)}  P_{\delta \delta}  \left( k,\chi \right) \,.
\ee
In the weak lensing approximation the magnification $\mu$ equals twice the convergence $\kappa$, where the latter equals the shear $\gamma$ at the level of two-point statistics, hence we can express $\cmm=4 \,C^{\kappa\kappa} = 4 \, \css$ \citep[see e.g.,][]{bas01} and subsequently  
\be
\label{eq:cmm}
\cmm (l) =9 \left( \frac{H_0}{c} \right)^4 \om^2 \int_0^{\chi_h} \d \chi \, \frac{g^2(\chi)}{a^2(\chi)}  P_{\delta \delta}  \left(k,\chi \right) \,.
\ee
In case of the angular galaxy number density power spectrum the weight function reads $q_i=p_i(\chi)\, b$  and subsequently we obtain  
\be
\label{eq:cpp}
\cpp (l) =\int_0^{\chi_h} \d \chi \, \frac{p^2 (\chi)}{\chi^2} \, b^2 \,P_{\delta \delta} \left( k,\chi \right) \,. 
\ee
Note that we have not yet specified the bias parameter $b$ and its functional dependence on $k$ and $z$; we address this further in Sect. \ref{sec:like}.
\subsection{Projected cross correlation power spectra}
\label{sec:cross-power}
Next we consider the cross correlation power spectra between our observables starting with the shear-magnification power spectrum 
\be
\label{eq:csm}
\csm (l) =\frac{9}{2} \left( \frac{H_0}{c} \right)^4 \om^2 \int_0^{\chi_h} \d \chi \, \frac{g^2(\chi)}{a^2(\chi)}  P_{\delta \delta}  \left( k,\chi \right) \,.
\ee
The corresponding relation for the shear-galaxy position power spectrum reads
\be
\label{eq:csp}
\csp (l)= \frac{3}{2}  \left( \frac{H_0}{c} \right)^2 \, \om \, \int_0^{\chi_h} \d \chi \, \frac{g(\chi) \, p (\chi)}{a(\chi) \, \chi} \,b \, r \, P_{\delta \delta} \left(k,\chi \right)\,, 
\ee
where $r$ denotes the correlation parameter for which, similar to the bias, we postpone an exact description to Sect. \ref{sec:like}.\\
Finally, we obtain
\be
\label{eq:cmg}
\cmp (l)=  \frac{3 \,H_0^2}{c^2} \, \om \, \int_0^{\chi_h} \d \chi \, \frac{g(\chi) \, p (\chi)}{a(\chi) \, \chi} \,b \, r \, P_{\delta \delta} \left(k,\chi \right)\,
\ee
as the expression for the magnification - galaxy position power spectrum.\\
We note the following interesting relations, which occur as a consequence of the polar-scalar filter functions $J_2$
\bea
\label{eq:rel1}
\cmm (l) &=& 4 \, \css (l) \quad = \quad 2 \, \csm (l) \\
\label{eq:rel12pt}
\ximm (\vt) &=& 4 \, \xip (\vt) \, \quad \neq \quad 2 \, \xism (\vt) 
\eea
and 
\bea
\label{eq:rel2}
\cmp (l) &=& 2 \, \csp (l)  \,\\
\label{eq:rel22pt}
\ximp (\vt) &\neq& 2 \, \xisp (\vt) \,.
\eea
We note that these relations can be used to create linear combinations that can asses the impact of systematics on individual probes. We will expand on this in Sect. \ref{sec:nulltests}.
\subsection{COSEBIs formalism}
\label{sec:cosebis}
The COSEBIs' formalism was developed in SEK10; we refer the reader to this paper for details beyond the brief summary presented in this section.\\
Throughout this paper we only consider filter functions that are logarithmic in $\vt$ as these filter functions comprise the second order shear information into significantly fewer COSEBI-modes compared to filter functions that are linear in $\vt$.

\subsubsection{Weak lensing}
\label{sec:coswl}
The COSEBIs shear E-mode, denoted as $E_n$, can be expressed as an integral over the shear 2PCF $\xi_{\pm}$ as
\be
\label{eq:eq1}
E^{\gamma \gamma}_n = \frac{1}{2} \int_\tmin^\tmax \d \vt \, \vt \, \left[\tss_{n+} (\vt) \, \xip (\vt)+ \tss_{n-} (\vt) \, \xim (\vt) \right] \,.
\ee
Note that for a properly constructed $\tss_{n+}$ as described below the corresponding $\tss_{n-}$ can be readily calculated as
\be
\label{eq:t-fromt+}
 \tss_{n-} (\vt)=  \tss_{n+} (\vt) +\int_0^\vt \d \theta \, \theta \, \tss_{n+} (\theta) \left( \frac{4}{\vartheta^2} - \frac{12 \theta^2}{\vartheta^4} \right)\,.
\ee
For further details on this the reader is referred to SEK10 and references therein. In the following we only describe the construction of $\tss_{n+}$.\\
In order to allow for a proper E/B-modes separation using a 2PCF over only a finite interval the filter functions $\tss_{n+}$ must meet the requirement 
\be
\label{eq:eq2}
\int \d \vt \, \vt\, \tss_{n+} (\vt) = 0 = \int \d \vt \, \vt^3 \, \tss_{n+} (\vt) \,.
\ee
In addition, the set of filter functions $\tss_{n+}$ must be orthonormal, i.e.
\be
\label{eq:eq3}
\frac{1}{\Delta \vt} \int_\tmin^\tmax \d \vt \, \tss_{n+} (\vt) \, \tss_{m+} (\vt) =\delta_{mn} \,.
\ee
The explicit construction of the logarithmic $T_{n+}$ is described in SEK10. The main steps of the construction are: 
\begin{itemize} 
\item A variable transformation $\vt \rightarrow z=\ln (\vt/\tmin)$ 
\item Expressing Eqs. (\ref{eq:eq2}, \ref{eq:eq3}) in $z$ with $\tss_{n+} (\vt) \rightarrow t^{\gamma \gamma}_{n+} (z)$
\item Expanding each $t^{\gamma \gamma}_{n+} (z)= \sum_{j=0}^{n+1} c_{nj} z^n$ 
\item Calculating the coefficients $c_{nj}$ from the conditions (\ref{eq:eq2}, \ref{eq:eq3})
\end{itemize}
Given $n$, the filter function $\tss_{n+}$ will be of order $n+1$ in $z$ as it needs to fulfill $n+1$ constraints, i.e. it must fulfill Eq. (\ref{eq:eq3}) for all $\tss_{m+}$ with $m \leq n-1$ and additionally it has to meet the two EB-mode separation constraints in Eq. (\ref{eq:eq2}). This implies that $\tss_{1+}$ is of order two. 

\subsubsection{Extension to clustering and magnification}
\label{sec:cosext}
Having determined the weak lensing COSEBIs filter functions $\tss_{n+}$, we can calculate the other five probe's COSEBIs similar to Eq. (\ref{eq:eq1})
\be
\label{eq:e2pt}
E^{AB}_n = \int_\tmin^\tmax \d \vt \, \vt \, T^{AB}_n (\vt) \, \xi^{AB} (\vt) \,.
\ee
In this paper we assume the same configuration space filter function $\tss_{n+}$ for all other probes, henceforth neglecting the superscripts.\\ 
In addition to Eq. (\ref{eq:e2pt}) the COSEBIs can be calculated directly from the power spectrum  
\be
\label{eq:epower}
E^{AB}_n = \frac{1}{2 \pi} \int \d l \, l \, W^{AB}_n (l) \, C^{AB} (l) \,,
\ee
which we only use as a consistency check since computing $C^{AB} (l) \rightarrow \xi^{AB} (\vartheta)$ using a fast Hankel-transformation and subsequently carrying out the finite integration in Eq. (\ref{eq:e2pt}) is significantly faster.\\
The Fourier filter functions $W^{AB}_n (l)$ are needed however for the computation of the COSEBI's covariance; they can be obtained from the $T_{n}$ as
\bea
\label{eq:wn1}
\wss_n (l) = \wmm_n (l)= \wpp_n (l) =  \wmp_n (l) =\int \d \vt \, \vt \, T_{n} (\vt) \, J_0 (l \vt) \,, \\
\wsp_{n}(l) = \wsm_{n}(l) = \int \d \vt \, \vt \, T_{n} (\vt) \, J_2 (l \vt) \,,
\eea
where the $J_{0/2}$ are a consequence of Eqs. (\ref{eq:2PCFshear} - \ref{eq:2PCFmp}).\\
\section{Modeling of Covariances}
\label{sec:cov}
\begin{figure}
\includegraphics[width=9cm]{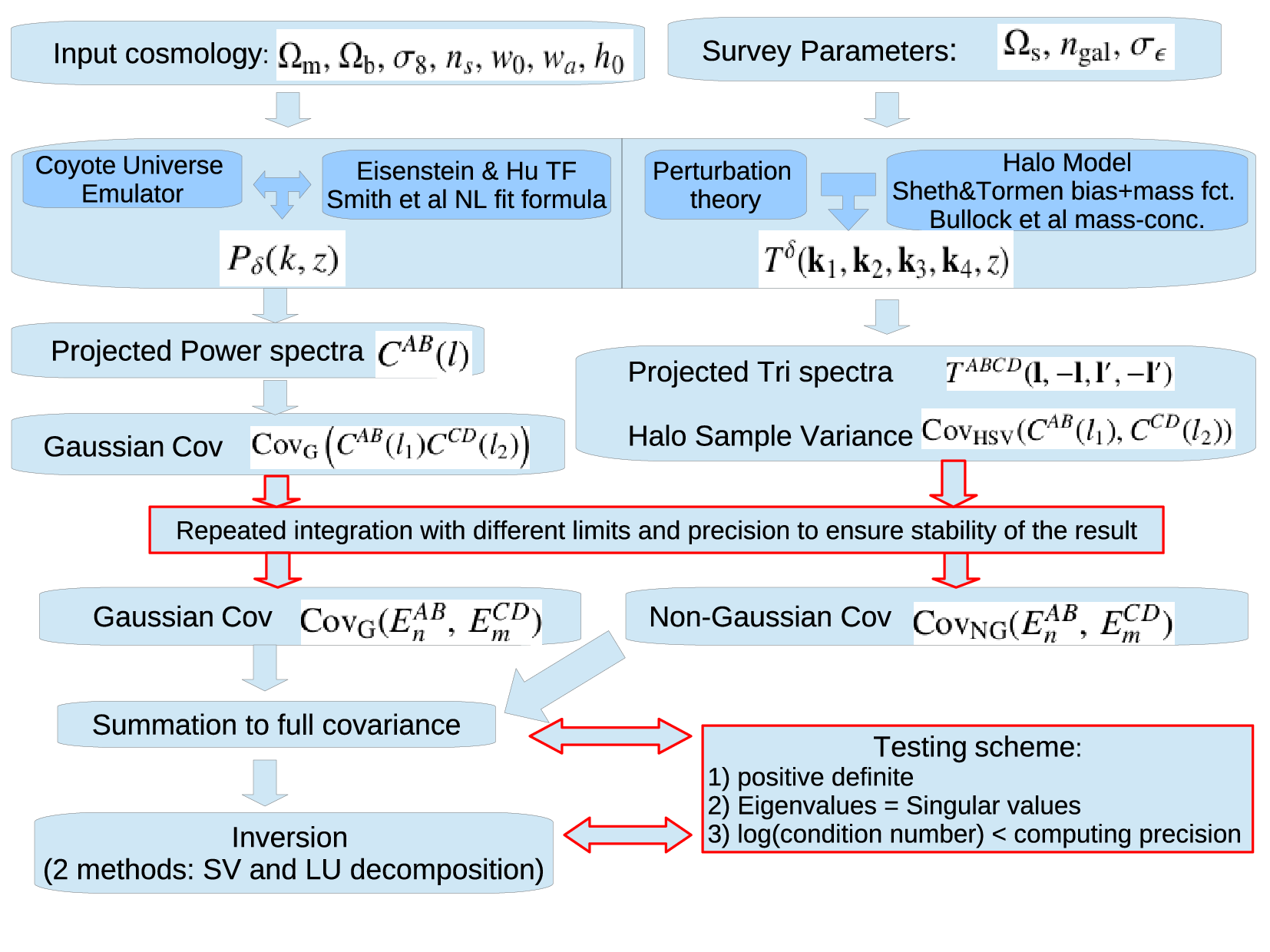}
 \caption{Schematic illustration of the modeling of the joint COSEBIs covariance for a given cosmology.}
         \label{fi:covflow}
\end{figure}

In this section we describe the covariance module of \textsc{CosmoLike v1.0} (see Fig. \ref{fi:covflow}). We start with explaining the modeling of covariances for projected power spectra; the expression for computing the COSEBIs covariance from the power spectrum covariance is straightforward, however the actual computation is easily affected by numerical uncertainties. We outline our method and cross checks at the end of this section.   
\subsection{Power spectrum covariances}
\label{sec:powcov}
\begin{figure}
\includegraphics[width=9cm]{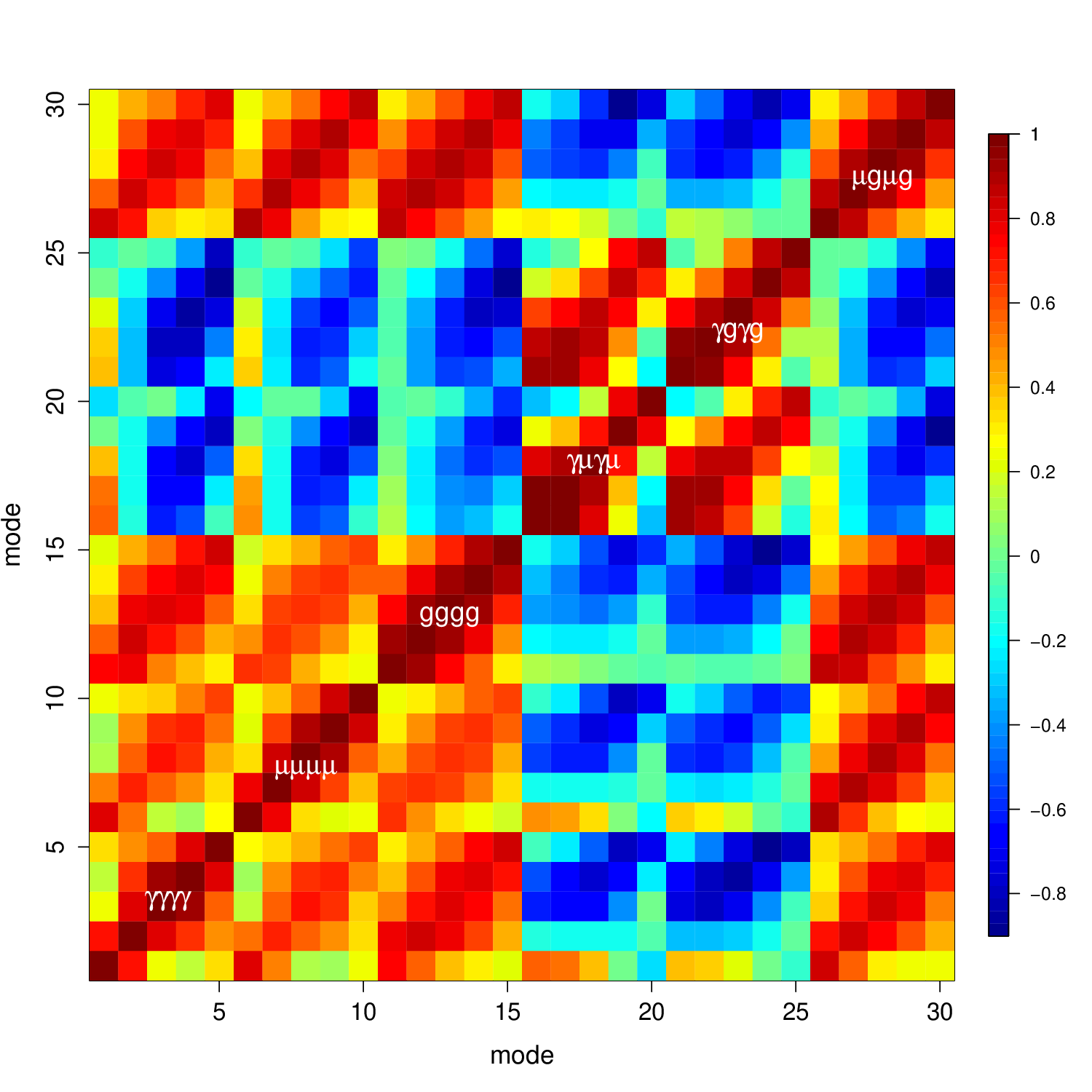}
 \caption{Full (Gaussian+Non-Gaussian) COSEBIs correlation matrix. Since we assume five modes for each of the six probes our data vector contains 30 data points, hence the covariance is 30 $\times$ 30. We indicate the corresponding auto-covariance block matrices in the plots.}
         \label{fi:covcor}
\end{figure}
Under the assumption that the density field is Gaussian (which means that the four-point function can be expressed in terms of two-point functions) the covariance of projected power spectra can be expressed as \citep{huj04} 
\bea
\label{eq:covhujain}
\mr{Cov_G} \left( C^{AB} (l_1) C^{CD} (l_2) \right) 
&=& \langle \Delta C^{AB} (l_1) \, \Delta C^{CD} (l_2) \rangle \nn \\
=  \frac{2 \pi \,\delta_{l_1,l_2}}{\Omega_\mr s l_1 \Delta l_1} && \hspace{-0.6cm} \left[(C^{AC}(l_1) +N^{AC}) (C^{BD}(l_1)+N^{BD}) \right. \nn \\
    &&  \hspace{-1.5cm}+ \left. (C^{AD}(l_1)+N^{AD}) (C^{BC}(l_1)+N^{BC}) \right]\,,
\eea
where the superscripts are to be replaced with $\gamma, g, \mu$ depending on the probe under consideration and $\Omega_\mr s$ denotes the survey volume. The covariance gets contributions from the signal $C(l)$ and a noise term $N$. Note that 
\be
N_{\gamma \gamma} = \frac{\sigma_\eps^2}{2 n_\mr{gal}} \;, \; N_{\mu \mu} = \frac{\sigma_\mu^2}{n_\mr{gal}}\;,\;N_{g g} = \frac{1}{n_\mr{gal}}\,, \\
\ee 
and all other noise terms are zero. We assume the intrinsic shape noise, $\sigma_\eps^2=0.32$, and note that the factor ``2'' in the denominator results from the fact that the shear has two components. For the magnification noise parameter we follow the arguments in \cite{kcd13} defining $\sigma_\mu=2 \, \sigma_\kappa/f^{1/2}$ for scaling relation based estimators of magnification \citep{hug11}, where $f$ denotes the fraction of galaxies for which magnification is measured and $\sigma_\kappa$ being the scatter of the convergence estimator. Being optimistic that the method in \cite{hug11} can be extended to late type galaxies we assume $\sigma_\mu=1.2$, noting that the uncertainty of this noise level is large.  \\
Since non-linear structure growth at late time induces significant non-Gaussianities in the density field, Eq. (\ref{eq:covhujain}) underestimates the error on cosmological parameters and needs to be amended by an additional term, i.e. $\mr{Cov}=\mr{Cov_G}+\mr{Cov_{NG}}$.
We model non-Gaussian covariance as the sum of the trispectrum contributions \citep{CH01,taj09}, including a sample variance term which describes the scatter in power spectrum measurements due to large scale density modes \citep{TB07, sht09},
\bea
\label{eq:covNG}
 \mr{Cov_{NG}}(C^{AB}(l_1),C^{CD}(l_2)) &=&  \frac{1}{\Omega_{\mr s}}\int_{|\mathbf l|\in l_1}\frac{d^2\mathbf l}{A(l_1)} \nn \\
 &\,&  \times \int_{|\mathbf l'|\in l_2}\frac{d^2\mathbf l'}{A(l_2)} T^{ABCD}(\mathbf l,-\mathbf l,\mathbf l',-\mathbf l') \nn \\
\eea
with $T^{ABCD}(\mathbf l,-\mathbf l,\mathbf l',-\mathbf l')$ defined as
\bea
\label{eq:tri2}
T^{\gamma^\alpha \mu^{\beta} g^{4-\alpha-\beta}} (l_1,l_2) &=& 2^\beta \, \left( \frac{3}{2} \frac{H_0^2}{c^2} \om \right)^{\alpha+\beta} \int_0^{\chi_h} \d \chi \, \left( \frac{g(\chi) \, \chi}{a(\chi)}\right)^{\alpha+\beta} \nn \\
&\times& \left( p(\chi) \, b \right)^{4-\alpha-\beta} \, \chi^{-6} \, \tdddd  \left( \frac{l_1}{\chi}, \frac{l_2}{\chi}, \chi \right) \,,
\eea
where we assume the correlation parameter $r=1$ and $\alpha, \beta \in [0;4]$. For example the pure shear tri-spectrum $\tssss$, and the pure galaxy position tri-spectrum $\tpppp$ read
\bea
\label{eq:tri1}
T^{\gamma^4} (l_1,l_2) &=& \left( \frac{3}{2} \frac{H_0^2}{c^2} \om \right)^4 \int_0^{\chi_h} \d \chi \, \frac{g^4(\chi)}{a^4(\chi) \, \chi^2} \; \tdddd  \left( \frac{l_1}{\chi}, \frac{l_2}{\chi}, \chi \right) , \\
T^{g^4} (l_1,l_2) &=&  \int_0^{\chi_h} \d \chi \, \frac{p^4(\chi)}{\chi^6} \, \tdddd \left( \frac{l_1}{\chi}, \frac{l_2}{\chi}, \chi \right) \,, 
\eea
respectively.

\subsection{Halo Model Trispectrum}
\label{ap:T}
We model the dark matter trispectrum using the halo model \citep{Seljak00, CS02}, which assumes that all matter is bound in virialized structures that are modeled as biased tracers of the density field. Within this model the statistics of the density field can be described by the dark matter distribution within halos on small scales, and is dominated by the clustering properties of halos and their abundance on large scales. In this model, the trispectrum splits into five terms describing the 4-point correlation within one halo (the \emph{one-halo} term $T^{\mr{1h}}$), between 2 to 4 halos (\emph{two-, three-, four-halo} term), and a so-called halo sample variance term $T^{\mr{HSV}}$, caused by fluctuations in the number of massive halos within the survey area,
\be
\label{eq:t}
T = T_{\mr{1h}}+\left(T_{\mr{2h},(22)}+T_{\mr{2h},(13)}\right)+T_{\mr{3h}}+T_{\mr{4h}}+T^{\mr{HSV}}\;.
\ee
The \emph{two-halo} term is split into two parts, representing correlations between two or three points in the first halo and two or one point in the second halo. As halos are the building blocks of the density field in the halo approach, we need to choose models for their internal structure, abundance and clustering in order to build a model for the trispectrum. Our implementation of the one-, two- and four-halo term contributions to the matter trispectrum follows \cite{CH01}, and we neglect the three-halo term as it is subdominant compared to the other terms at the scales of interest for this analysis. Specifically, we assume NFW halo profiles \citep{NFW} with the \citet{Bullock01} fitting formula for the halo mass--concentration relation $c(M,z)$, and the \citet{ST99} fit functions for the halo mass function $\frac{ dn}{dM}$ and linear halo bias $b(M)$, neglecting terms involving higher order halo biasing.

\subsection{COSEBIs covariances}
\label{sec:ngcov}
Following \citet{sht09} the halo sample variance term can be is calculated as
\bea
\nonumber \mr{Cov_{HSV}}\hspace{-0.4cm} &(& \hspace{-0.4cm} C^{AB}(l_1),C^{CD}(l_2)) \hspace{+0.4cm} = \hspace{+0.4cm}2^\beta \, \left( \frac{3}{2} \frac{H_0^2}{c^2} \om \right)^{\alpha+\beta} \nn \\
 &\times& \int_0^{\chi_\mr h} d\chi \left(\frac{d^2 V}{d\chi d\Omega}\right)^2  \left( \frac{g(\chi) \, \chi}{a(\chi)}\right)^{\alpha+\beta} \left( p(\chi) \, b \right)^{4-\alpha-\beta} \nn \\
  &\times &  \int d M \frac{d n}{d M} b(M)\left(\frac{M}{\bar{\rho}}\right)^2 |\tilde{u}(l_1/\chi, c(M,z(\chi))|^2 \nn  \\
 &\times & \int d M' \frac{d n}{d M'} b(M')\left(\frac{M'}{\bar{\rho}}\right)^2 |\tilde{u}(l_1/\chi, c(M',z(\chi))|^2 \nn \\
&\times& \int_0^\infty \frac{k dk}{2\pi}P_\delta^{\mr{lin}}(k,z(\chi))|\tilde W(k\chi \Theta_{\mr s})|^2 \,,
\eea
with $\tilde{u}(l_1/\chi, c(M,z(\chi))$ the normalized Fourier transform of the halo density profile.\\
Adding Gaussian (Eq. \ref{eq:covhujain}) and non Gaussian covariance (Eq. \ref{eq:covNG}) and subsequently integrating over $l_1$ and $l_2$ we obtain the final COSEBI's covariance
\bea
\label{eq:covnongaussian}
\mr{Cov} \left( E^{AB}_n, E^{CD}_m \right) &=& \frac{1}{4 \pi^2} \left[ \int_0^\infty \d l_1\;l_1\, W^{AB}_n (l_1)\, W^{CD}_m (l_1)\,  \mr{Cov}_\mr{G} (l_1) \right.  \nn \\
+ && \hspace{-0.9cm} \left. \int_0^\infty \d l_1 \, l_1 \, W^{AB}_n (l_1) \int_0^\infty \d l_2 \, l_2 \, W^{CD}_m (l_2) \mr{Cov}_\mr{NG} (l_1, l_2) \right] \,. \nn \\
\eea
This integration must be tested thoroughly for convergence and stability of the result with respect to numerical integration precision, upper and lower limit of the integration. If the result is stable we verify that the covariance and its inverse is positive definite. In Fig. \ref{fi:covcor} we show the correlation matrix of the full covariance matrix. Since the COSEBIs filter functions average over all Fourier modes/angular scales, it is difficult to have an intuitive understanding of this matrix. We show the impact on likelihood contours when neglecting the Non-Gaussian terms in Fig. \ref{fi:like2}.

\section{Likelihood Analysis}
\label{sec:like}
\begin{figure*}
\includegraphics[width=17cm]{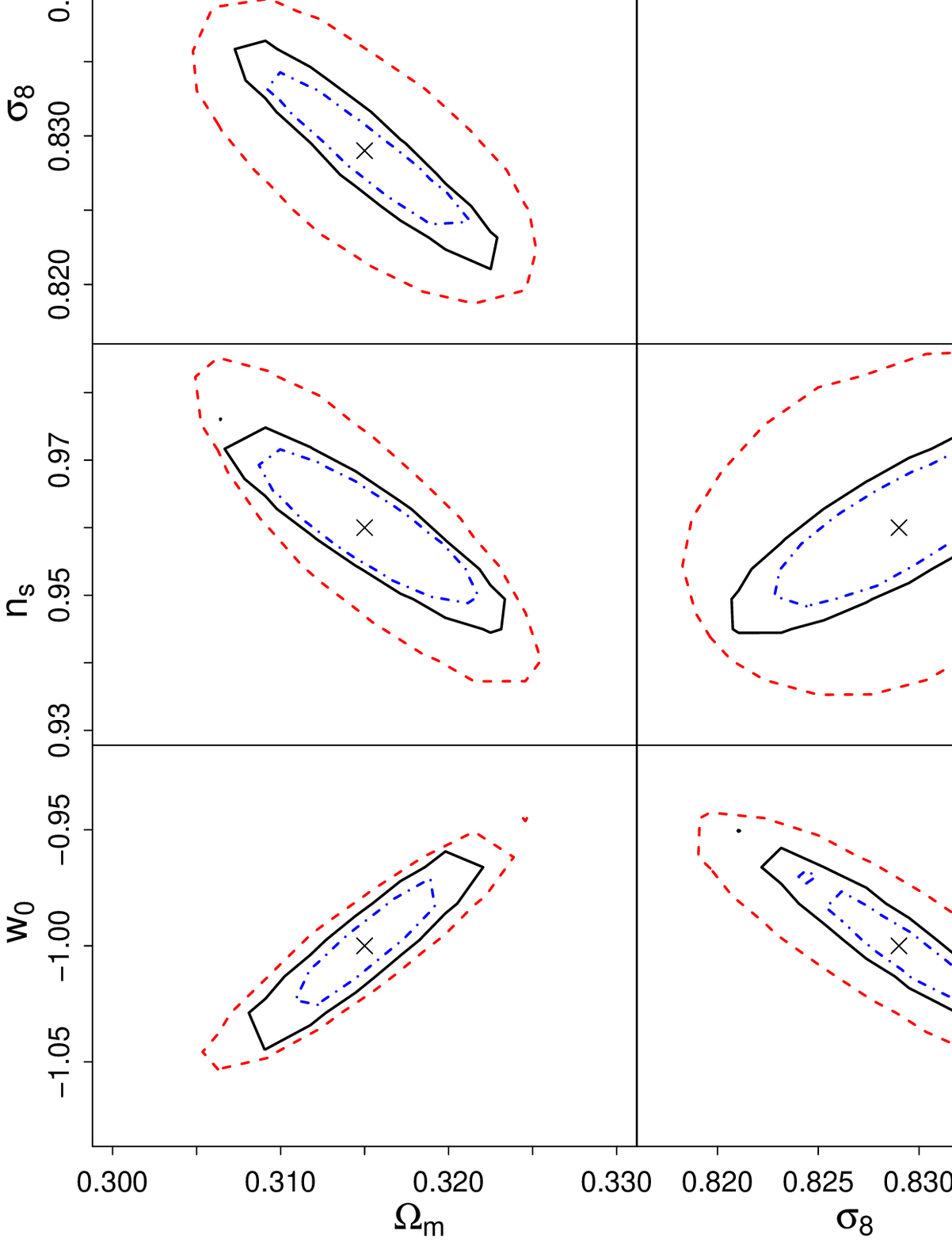}
 \caption{Likelihood analysis in four dimensional cosmological parameter space. We show the 68\% credible regions and marginalize over the all other parameters not shown in a given panel. We compare CP analyses with and without magnification (\ti{Black/solid} and \ti{red/dashed}, respectively). For the full six probe data vector $\vek E=(\ess$, $\emm$, $\epp$, $\esm$, $\esp$, $\emp)$) we also show the difference when using Gaussian instead of Non-Gaussian covariances (\ti{blue/dot-dashed} and \ti{black/solid}, respectively).}
         \label{fi:like2}
\end{figure*}

\begin{table}
\begin{center}
\caption{Cosmological parameter ranges used in the likelihood analyses.}
\label{tab:tab2}
\begin{tabular}{c c c}\hline
parameter  &  flat prior & fiducial \\ \hline
$\om$  &  $[0.05;0.8]$ & $0.315$\\
$\sig$ &  $[0.4;1.2]$ & $0.829$\\
$w_0$  &  $[-1.8;-0.2]$ & $-1.0$\\
$n_s$  &  $[0.6;1.2]$ & $0.96$\\
$w_a$  &   $[-2.0;2.0]$ & $0.0$ \\

\end{tabular}
\end{center}
\end{table}
\begin{figure*}
\includegraphics[width=17cm]{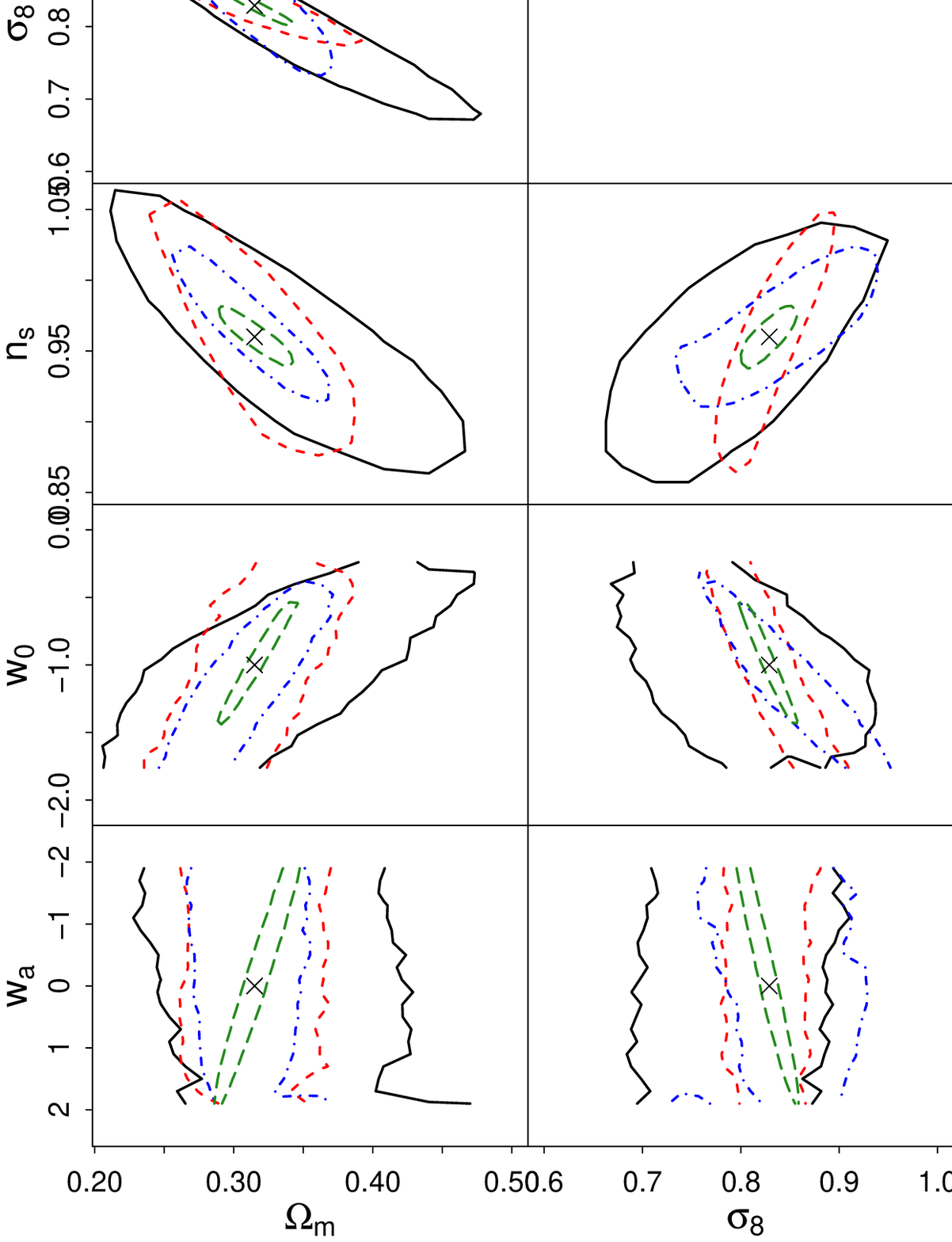}
 \caption{Likelihood analysis in five-dimensional cosmological parameter space as described in the text. We show the 68\% credible regions for four different likelihood analyses, i.e. individual probes of cosmic shear, galaxy-galaxy lensing and galaxy clustering, compared to a joint analysis of all three probes (see legend for details).}
\label{fi:like1}
\end{figure*}
\textsc{CosmoLike} samples the parameter space using parallel MCMC \citep{gow10} implemented through the \textsc{emcee} python package\footnote{http://dan.iel.fm/emcee/}. The MCMCs presented in this paper consist of at least 200000 steps and have been checked for convergence. In the following we simulate various likelihood analyses using the data vector in Eq. (\ref{eq:dv}) and covariance in Eq. (\ref{eq:covconcept}) or subsets thereof for survey parameters close to what is expected for the Dark Energy Survey ($\Omega_\mr s=5000 \, \mr{deg^2}, n_\mr{gal}=10/\mr{arcmin}^2$). The range of the cosmological parameter space considered in the different analyses is summarized in Table \ref{tab:tab2}; the fiducial cosmology is similar to the Planck+WMAP polarization best-fit results \citep{planckcosmo13}. We calculate the likelihood as described in Eq. (\ref{eq:likelihood}) and subsequently the posterior probability via Eq. (\ref{eq:bayes}).  The contour plots in Figs. \ref{fi:like1} -\ref{fi:like3} show the marginalized probability calculated as
\be
\label{eq:marg}
L (\vek E| \pco (2D) ) = \int_{\pco'(min)}^{\pco'(max)} \d \pco'  \, L (\vek E|\pco') \,, 
\ee
where $\pco'$ denotes the remaining cosmological parameters when subtracting the considered two parameters from the full parameter set.\\
The schematic illustrations in Fig. \ref{fi:modelscheme} and Fig. \ref{fi:covflow} show the computation of the model vector and covariance, respectively. Note that the COSEBIs model vector is computed from the corresponding correlation function over an interval of $[1';400']$. Computing time for the full COSEBIs model vector is $<$1 sec per cosmology.

\paragraph*{Comparing individual and combined probes}
In our first likelihood analysis we compare the cosmological information of individual probes to a CP analysis. Figure \ref{fi:like1} shows constraints from the single probe data vectors $\ess$, $\esp$, and $\epp$ (black/solid, red/dashed, and blue/dash-dotted, respectively) and their corresponding covariance matrices (submatrices of Eq. \ref{eq:covconcept}) to the CP data vector that consists of all three probes $\vek E=(\ess$, $\esp$, $\epp)$ (green/long-dashed contours). We consider a five-dimensional cosmological parameter space ($\om$, $\sig$, $n_s$, $w_0$, and $w_a$) free of nuisance parameters (see Sect. \ref{sec:bias} for bias modeling uncertainties).\\
From Fig. \ref{fi:like1} it is clear (and expected) that our non-tomographic likelihood analysis has little constraining power for time dependent dark energy models if one considers the probes individually. For the CP analysis, the constraints on $\om$, $\sig$, $n_s$ are improved substantially, which allowing us to put tight constraints on the combination of $w_0$-$w_a$. We expect these constraints to significantly improve if tomographic information is included. 

\paragraph*{Adding Magnification}
We further extend the analysis by adding the magnification auto and cross correlation probes to the CP data vector, which means we add three new second-order statistics to the existing three. For this part of the analysis we consider a four dimensional parameter space only, more precisely we fix $w_a=0$.  
Figure  \ref{fi:like2} compares two different data vectors, namely $\vek E=(\ess, \emm, \epp, \esm, \esp, \emp$) (black/solid and blue/dot-dashed contours) and $\vek E=(\ess, \epp, \esp)$ (red/dashed contours); for the first case we further compare analyses using the full Non-Gaussian covariance (black) to using the Gaussian approximation (blue).\\ 
We find a clear improvement in cosmological information when including the magnification auto and cross probes in the data vector compared to using shear and clustering only. Although the inclusion of magnification doubles the number or probes in the analysis, the increase in information is not expected to be larger due to the large degeneracy of shear and magnification as a cosmological probe. The improvement will likely be more substantial when including nuisance parameters to account for uncertainties in shear calibration, photo-z, intrinsic alignment, and baryons. These uncertainties affect both probes differently, hence adding magnification to the CP framework is a valuable resource of information to mitigate the impact on parameter constraints.
\paragraph*{Comparing Gaussian and Non-Gaussian covariances}
In Fig. \ref{fi:like2} we also show the difference in parameter constraints when using Gaussian instead of Non-Gaussian covariances for the six probe CP vector. For likelihood analyses of individual probes such comparisons have been carried out in previous papers \citep[see e.g.][for cosmic shear]{taj09,esh09}, this however is the first time that such a comparison 1) is shown for the CP case and 2) includes the Halo Sample Variance term in the covariance. We find that there is a clear difference in parameter constraints when neglecting the higher order correlations of the density field in the computation of the covariance, which indicates that the precise modeling of these higher order terms is non-negligible for accurate parameter constraints.  

\subsection{Uncertainties from Bias and Correlation Parameters}
\label{sec:bias}
\begin{figure*}
\includegraphics[width=17cm]{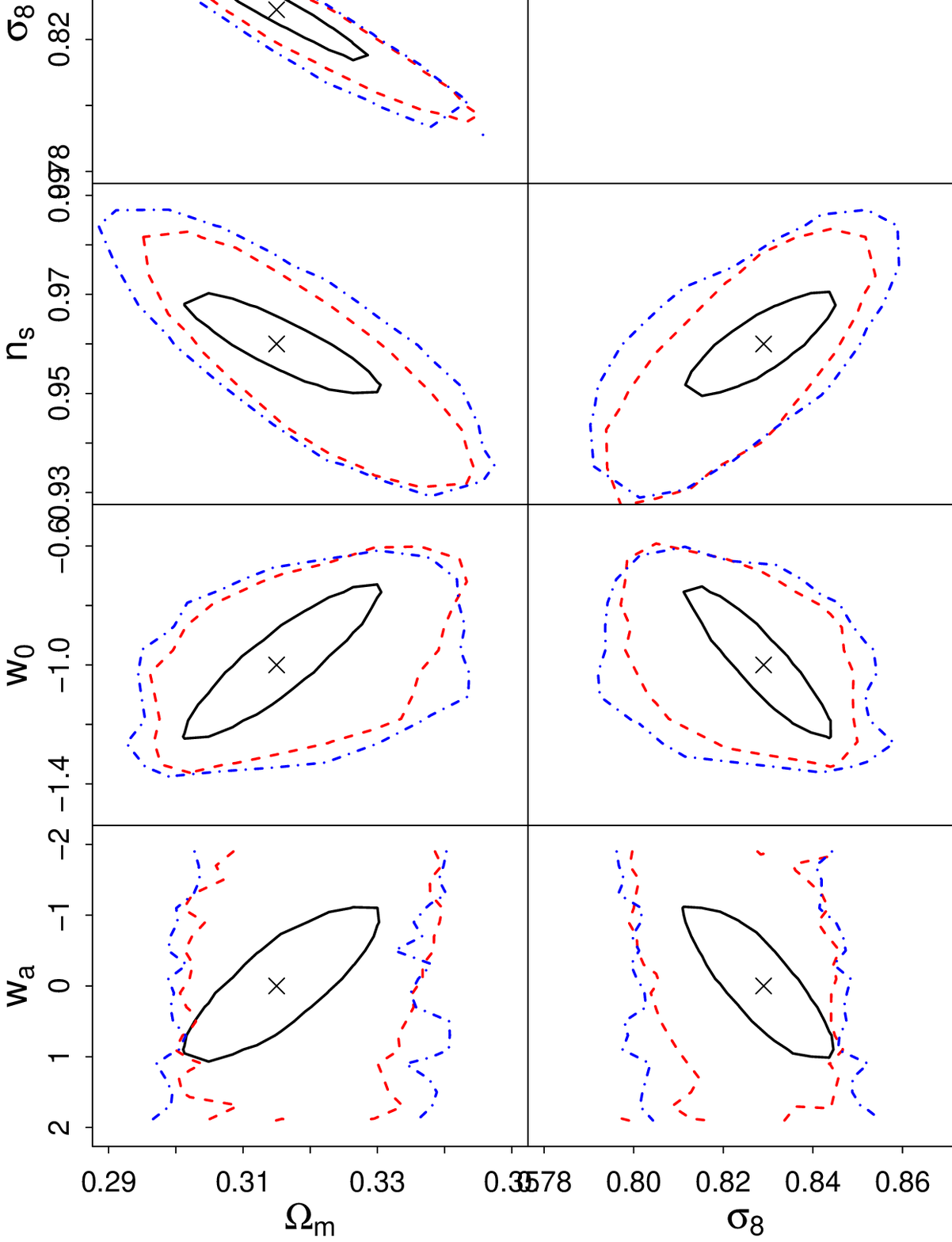}
 \caption{Likelihood analysis in five-dimensional cosmological parameter space using the $\ess$, $\emm$, $\epp$, $\esm$, $\esp$, $\emp$ data vector and the corresponding Non-Gaussian covariance. We show the 68\% credible regions where \ti{Black/solid} contours correspond to a likelihood analysis assuming perfect knowledge of bias and \ti{red/dashed} and \ti{blue/dot-dashed} contours correspond to marginalizing over two different bias modeling scenarios (see text for further details).}
         \label{fi:like3}
\end{figure*}

Understanding the relation of galaxies and their dark matter environment is an important aspect of any cosmological parameter estimation that includes clustering information. Constraining and modeling this relation is an active field of research in theory \citep[e.g.,][]{zbw05,mcr09} and observations \citep[e.g.,][]{clb12,msb12,jrk12}. In practice, any bias model will have to be finetuned to the considered data set (galaxy population/morphology and redshift distribution). Guidance on any parametrization from first physical principles is limited; measurements rely mostly on configuration space quantities, i.e. a parametrization in $r,z$ or $\vartheta$.\\
Bias and correlation parameter can also be parametrized \citep[e.g.][]{bas01, ber09} as a function of $k,z$ in Eqs. (\ref{eq:cpp}, \ref{eq:csp}), more precisely 
\be
\label{eq:bias1}
b^2 (|\vek k|, \chi) = \frac{P_{gg} (|\vek k|, \chi)}{P_{\delta \delta} (|\vek k|, \chi)} \,,
\ee
and
\be
\label{eq:bias2}
r (|\vek k|, \chi) = \frac{P_{\delta g} (|\vek k|, \chi)}{\sqrt{P_{\delta \delta} (|\vek k|, \chi) P_{g g} (|\vek k|, \chi)}}\,.
\ee
with $P_{g g}$ being the observable galaxy number density power spectrum and $P_{\delta g}$ being the galaxy-dark matter cross power spectrum. In contrast, we do not parametrize $b$ and $r$ in 3D Fourier space, but directly as a function of the quantity that enters the likelihood analysis, i.e. we define the relation between the observable galaxy number density COSEBIs $E^{gg}_n$ and the projected dark matter density COSEBIs $E^\mr{mm}_n$ as
\be
\label{eq:cosbias1}
E^{gg}_n= X_n \, E^\mr{mm}_n \,,
\ee
and similarly 
\be
\label{eq:cosbias2}
E^{\gamma g}_n= Y_n \, E^{\gamma \mr m}_n \,.
\ee
The parameters $X_n, Y_n$ express the relation between projected dark matter and galaxy number density; their range of uncertainty reflects the uncertainty in modeling $b$ and $r$ as a function of redshift and scale. COSEBIs mix angular scales and we do not consider a tomographic analysis our uncertainty in  describe the effective uncertainty of $b$ and $r$ averaged over a large mix of scales and redshift.  \\ 
For the likelihood analysis results shown in Fig. \ref{fi:like3} we assume that all $X_n$ are uncorrelated; the same holds for $Y_n$ and also for combinations of $Y_n$ and $X_n$. We introduce ten (nuisance) parameters (five $X_n$ and $Y_n$) to model the uncertainty between dark and luminous matter and allow them to vary independently. More precisely, we model $X_n=(b_\mr{fid}+\Delta b_n)^2$ and $Y_n=(b_\mr{fid}+\Delta b_n)(r_\mr{fid}+\Delta r_n)$, where $\Delta b_n$ and $\Delta r_n$ are drawn from a Gaussian probability distribution with $\sigma^2=0.1$ for a more optimistic and $\sigma^2=0.2$ for a more pessimistic scenario (labeled scenario 1 and 2, respectively). Our method can be interpreted as ``self-calibration'' with Gaussian priors centered around the fiducial values of bias and correlation parameter ($b_\mr{fid}=1.2$ and $r_\mr{fid}=1.0$). We address the importance of the prior below.\\
The simulated likelihood analysis in Figure \ref{fi:like3} shows the results in $\om$, $\sig$, $w_0$, $n_s$, $w_a$ parameter space assuming perfect knowledge of bias and correlation parameter (black/solid contours), and using the parametrization in Eqs. (\ref{eq:cosbias1}, \ref{eq:cosbias2}) for various scenarios. As expected, marginalizing over uncertainties in $X_n$ and $Y_n$ significantly weakens cosmological constraints across all parameters and the effect is slightly stronger for the pessimistic bias scenario. The small difference between optimistic and pessimistic scenario indicates that our priors on $\Delta b$ and $\Delta r$ hardly affect the self-calibration procedure, i.e. choosing a larger $\sigma^2$ will not significantly alter the parameter constraints. \\ 
We note that our bias parametrization is conservative since the scale and redshift dependence of $b$ and $r$ induces correlations in $X_n$ and $Y_n$; including these correlations decreases the nuisance parameter range that we marginalize over. We point out that ideally the inclusion of clustering information on small scales requires sophisticated HOD modeling and marginalization over the corresponding HOD parameters. Our self-calibration method can be seen as a lower bound; more information on galaxy formation implemented via HOD modeling can only improve constraints.\\ 
In the following we suggest a measurement framework to make progress on $X_n$ and $Y_n$ and their range of uncertainty observationally. In the context of the aperture mass dispersion this method has been suggested in \cite{wae98} to detect scale dependence of galaxy bias by combining second-order statistics of shear and clustering \citep[see][for application to data]{hwg02,clb12}.\\
For the COSEBIs the corresponding relations read
\be
\label{eq:Xbias}
X_n= f_X \,  \frac{E^{gg}_n}{E^{\gamma \gamma}_n} \,
\ee
and
\be
\label{eq:ybias}
Y_n= f_Y \,  \frac{E^{\gamma g}_n}{\sqrt{E^{\gamma \gamma}_n \, E^{g g}_n}} \, .
\ee
The functions $f_X$ and $f_y$ depend weakly on cosmology \citep{hwg02,skw06}, which can be mitigated even further by employing strong priors from independent experiments, e.g. Planck. This method allows to measure and constrain a ``mode-dependent'' bias for the COSBEBIs. We however note that on cosmological scales galaxy bias has little scale but significant redshift dependence and that this method should be applied within sufficiently small tomography bins. \\
Analogous relations can be derived using $E^{\mu \mu}_n$ instead of $E^{\gamma \gamma}_n$ and/or using $E^{\mu g}_n$ instead of $E^{\gamma g}_n$. We emphasize that magnification can provide additional information to constrain the relation between dark and luminous matter; at the very least it provides a valuable cross check/nulltest for the above method.\\ 

\subsection{Nulltests involving shear and magnification}
\label{sec:nulltests}
The fact that including magnification contributes only little to the cosmological constraints is not unexpected given the degeneracy of shear and magnification. This degeneracy however allows us to test for and to quantify systematics that affect both probes differently. As an example we will assume that one of the most important contaminations of cosmic shear, i.e. intrinsic alignment, does not affect magnification. We can express the observed shear power spectrum as a the sum of the true shear power spectrum and the two intrinsic alignment components II (correlation of the intrinsic ellipticity with the local density field) and GI (correlation of foreground galaxy ellipticity with background shear) \citep{his04,ber09,job10}
\bea
\label{eq:IA1}
\css_\mr{obs} (l) &=& \css (l) + C^\mr{II} (l)+ C^\mr{GI}  (l) \,, \\
\label{eq:IA2}
\cmm_\mr{obs} (l) &=& \cmm (l)\,.
\eea
Using Eq. (\ref{eq:rel12pt}) in terms of the COSEBIs we can rewrite Eqs. (\ref{eq:IA1}, \ref{eq:IA2}) as
\be
\label{eq:IA4}
 E^\mr{II}_n + E^\mr{GI}_n =E^{\mu \mu}_n (obs) \, - \, 4  \, E^{\gamma \gamma}_n (obs)  \,,
\ee
thereby constraining intrinsic alignment. We note that contaminations similar to IA for shear might exist for magnification as well; correlations of the local density field with the intrinsic size of the galaxies ($\mr{II}^\mu$), and/or correlations of the foreground galaxy size with the magnification of a background galaxy ($\mr{GI}^\mu$) are likely. As discussed in \cite{slm12} magnification estimators which are not based on the excess of number densities but on size measurements have little correlation with their environment \citep[e.g.][]{cfn05,mag10} therefore potentially allowing the above technique to be successful.\\
In any case the above nulltest can be used for other types of contaminations which dominate the shear related quantity but do not affect magnification; the best example probably being shear calibration.\\
The number of nulltests is not limited to the magnification and shear auto-correlations but additional constraints can be gained from three other relations similar to Eqs. (\ref{eq:rel1}, \ref{eq:rel2}). As a prerequisite we define new COSEBIs filter functions for $\gamma g$ and $\gamma \mu$, denoted as $T_{n}'$ such that 
\bea
\label{eq:wn3}
\wss_n (l) &=& \wsp_{n}(l) = \wsm_{n}(l) \,, \\
\int \d \vt \, \vt \, T_{n} (\vt) \, J_0 (l \vt) &=& \int \d \vt \, \vt \, T_{n}' (\vt) \, J_2 (l \vt) \,,
\eea
with $T_{n}$ still being the original filter function $\tss_{n+}$ defined in Sect. \ref{sec:coswl}. Calculating the new COSEBIs $E^{\gamma g}_n$ and $E^{\gamma \mu}_n$ as an integral over the corresponding correlation functions $\xisp$ and $\xism$ using the $T_{n}'$ we derive the relations
\bea
E^{\mu \mu}_n &=& 2 \, E^{\gamma \mu} \,, \\
E^{\gamma \mu}  &=& 2 \, E^{\gamma \gamma}_n\,, \\
E^{\mu g}_n &=& 2 \, E^{\gamma g}_n \,, 
\eea
which do not hold for the correlation functions (see Eqs. \ref{eq:rel12pt}, \ref{eq:rel22pt}).\\
These relations or linear combinations thereof can be used to define nulltests and subsequently constrain astrophysical uncertainties. 

\section{Conclusions} 
\label{sec:conc}
In this paper we introduce \textsc{CosmoLike v1.0}, a coherent analysis framework to extract cosmological constraints from all second-order statistics that can be derived from a galaxy position ($g$), shear ($\gamma$), and magnification ($\mu$) catalog.\\
The \textsc{CosmoLike} prediction code module allows for fast modeling of multi-probe data vectors consisting of various second-order statistics. These are computed from density power spectra that are generated by the Coyote Universe Emulator or a modified Halofit implementation. The \textsc{CosmoLike} covariance module utilizes a halo model implementation to computing non-Gaussian covariances for all aforementioned projected quantities and their cross terms. We then generate a CP data vector from our prediction code assuming a fiducial cosmology and test this data vector (and subsets thereof) in several likelihood analyses, the most extensive one covering five cosmological dimensions $\om$, $\sig$, $w_0$, $n_s$, $w_a$ and a ten parameter self-calibration bias model.\\
The analysis scheme suggested in this paper differs from previous work in several ways: First, we include all second-order cross statistics of the observables ($g$, $\gamma$, $\mu$) into the data vector (thereby increasing the sources of cosmological information), second, we model all cross terms in the covariance matrix, and third, we include all higher order correlations of the density field in the covariance matrix. Furthermore, we employ the COSEBIs formalism to quantify the information content, thereby solving the cosmic shear E/B-mode problem and introducing a data compression scheme for the other five two-point statistics.\\ 
Not surprisingly, we find substantial improvement in parameter constraints when using the CP data vector instead of the individual probes, and a sizable increase in the CP likelihood contours when fitting for galaxy bias instead of assuming it perfectly known.\\
The most interesting results of this paper are the changes in likelihood contours when including magnification in the data vector and when modeling the higher order moments of the density field in the covariance matrix. \\
We find a noticeable improvement when including magnification and all cross probes in addition to cosmic shear, galaxy-galaxy lensing, and galaxy clustering. Although the inclusion of magnification doubles the number of second-order statistics in the data vector the strong degeneracy of information from magnification and shear and the large assumed noise level $\sigma_\mu=1.2$ prevent this improvement to be more significant. We note that magnification in contrast to shear and clustering is a relatively recent cosmological probe, hence the assumed noise level might be too pessimistic.\\
The degeneracy of shear and magnification allows for interesting constraints on systematics, e.g. intrinsic alignment, shear calibration errors, photo-z uncertainty, etc. We outline several relations that hold in the absence of these systematics and suggest extensions of these nulltests. We also emphasize that in the presence of nuisance parameters describing the uncertainty from these systematics, the information increase when including magnification will likely be more significant.\\ 
Regarding covariances we find that neglecting the higher order terms in their modeling leads to a clear underestimation of error bars. We emphasize that forecasting exercises for a CP analysis similar to ours should incorporate Non-Gaussian covariances including all cross terms of probes.\\
In the future we plan to extend \textsc{CosmoLike} to other second-order statistics whose distribution follow the same likelihood function, e.g. CMB polarization, CMB lensing, CMB temperature correlations, and of course also to tomography for the probes considered in this paper.    
It is straightforward to apply this analysis scheme to a any data set from which can measure projected correlation functions. Before extracting meaningful information from data however, the framework described here needs extensions. For example, implementing a detailed HOD-model approach \citep{bmc13}, adding the parametrization of nuisance parameters, such as  baryons, intrinsic alignment, photo-z calibration, and shear calibration is required. \\

\section*{Acknowledgments}
We thank Eric Huff, David Weinberg, Scott Dodelson, Bhuvnesh Jain, and Gary Bernstein for very useful discussions and advice. This paper is based upon work supported in part by the National Science Foundation under Grant No. 1066293 and the hospitality of the Aspen Center for Physics. The research of TE and EK was funded in part by NSF grant AST 0908027 and U. S. Department of Energy grant DE-FG02-95ER40893. The work was supported by the Deutsche Forschungsgemeinschaft with the program TR33 `The Dark Universe'.


\label{lastpage}
\end{document}